\documentclass{article}  
\usepackage{breckenridge}
\usepackage{graphicx,wrapfig,epsfig}
\usepackage[figuresright]{rotating}
\frompage{000} \topage{000}                                              
\def\rightmark{Global Observables at PHENIX}
\def\leftmark{A. Bazilevsky}
 
\title{Global Observables at PHENIX}
\authors{
{Alexander Bazilevsky$^1$, for the PHENIX Collaboration}\\[2.812mm]
{\normalsize
\hspace*{-8pt}$^1$ RIKEN BNL Research Center, Brookhaven National Laboratory,\\
Upton, NY, USA
}}
 
\abstract{
We present PHENIX recent results on charged particle 
and transverse energy densities  
measured at mid-rapidity in Au-Au
collisions at $\sqrt{s_{_{NN}}}$=130 GeV and 200 GeV 
over broad range of centralities. The mean transverse
energy per charged particle is derived. The first PHENIX measurements 
at $\sqrt{s_{_{NN}}}$=19.6 GeV are also presented.
A comparison with calculations from various
theoretical models is performed. 
}
 
\begin{document}
 
\maketitle
\setcounter{page}{1}

\section{Introduction}\label{intro}

Charged particle multiplicity ($N_{ch}$) and transverse energy ($E_{T}$) 
are global variables which give excellent 
characterization of the high energy nucleus-nucleus collisions, thus 
providing information about the initial conditions \cite{intr1}. 
They help to constrain the wide 
range of theoretical predictions and discriminate among various mechanisms of 
particle production. $N_{ch}$ measurements also provide one with an 
opportunity to study high density QCD effects in relativistic nuclear 
collisions \cite{kln}. 

First results for $N_{ch}$ and $E_{T}$ at mid-rapidity in Au-Au collisions 
at $\sqrt{s_{_{NN}}}=130$~GeV and 200~GeV measured with the PHENIX detector 
were published in \cite{prl_nch,prl_et,qm01_etnch,qm02_etnch}. 
In this paper we extend the $N_{ch}$ and $E_{T}$ analysis to more 
peripheral collisions, up to 65\%-70\% centrality class 
corresponding to the average number of participating nucleons 
$N_{p} \approx 22$. 
The results for the $<E_{T}>/<N_{ch}>$ are presented 
at $\sqrt{s_{_{NN}}}=19.6$~GeV, 130~GeV and 200~GeV. 

We used the same experimental techniques to analyze data obtained 
at different beam energies. The minimum bias (MB) trigger 
in $\sqrt{s_{_{NN}}}=19.6$~GeV data was formed by the 
coincidence of two beam-beam counters. In $\sqrt{s_{_{NN}}}=130$~GeV 
and 200~GeV data the MB trigger additionally required the signals from 
two zero degree calorimeters \cite{prl_nch,prl_et,qm01_etnch,qm02_etnch}. 
The corrections for the measured 
$E_{T}$ and $N_{ch}$ related to particle composition and mean 
transverse momentum are performed based on PHENIX results 
obtained at $\sqrt{s_{_{NN}}}=130$~GeV \cite{prl_h,prl_l} 
and 200 GeV \cite{qm02_h}. $\sqrt{s_{_{NN}}}=19.6$~GeV data 
corrections are based on NA49 measurements obtained 
at $\sqrt{s_{_{NN}}}=17.2$~GeV \cite{na49}.

The part of the PbSc electromagnetic calorimeter used for the $E_{T}$ 
measurements covers the pseudorapidity range $|\eta|\leq 0.38$
with an azimuthal aperture of $\Delta\phi=44.4^\circ$ in 
Run-1 data ($\sqrt{s_{_{NN}}}=130$~GeV) and $\Delta\phi=112^\circ$ in 
Run-2 data ($\sqrt{s_{_{NN}}}=19.6$~GeV and 200 GeV).
The pad chambers used for $N_{ch}$ measurements have a 
fiducial aperture of $|\eta|\leq 0.35$ and 
$\Delta\phi=90^\circ$ and $180^\circ$ in the Run-1 and Run-2 
data respectively.

\section{Results}  

\begin{figure}[t]
\begin{minipage}[t]{75mm}
\centerline{\includegraphics[height=75mm]{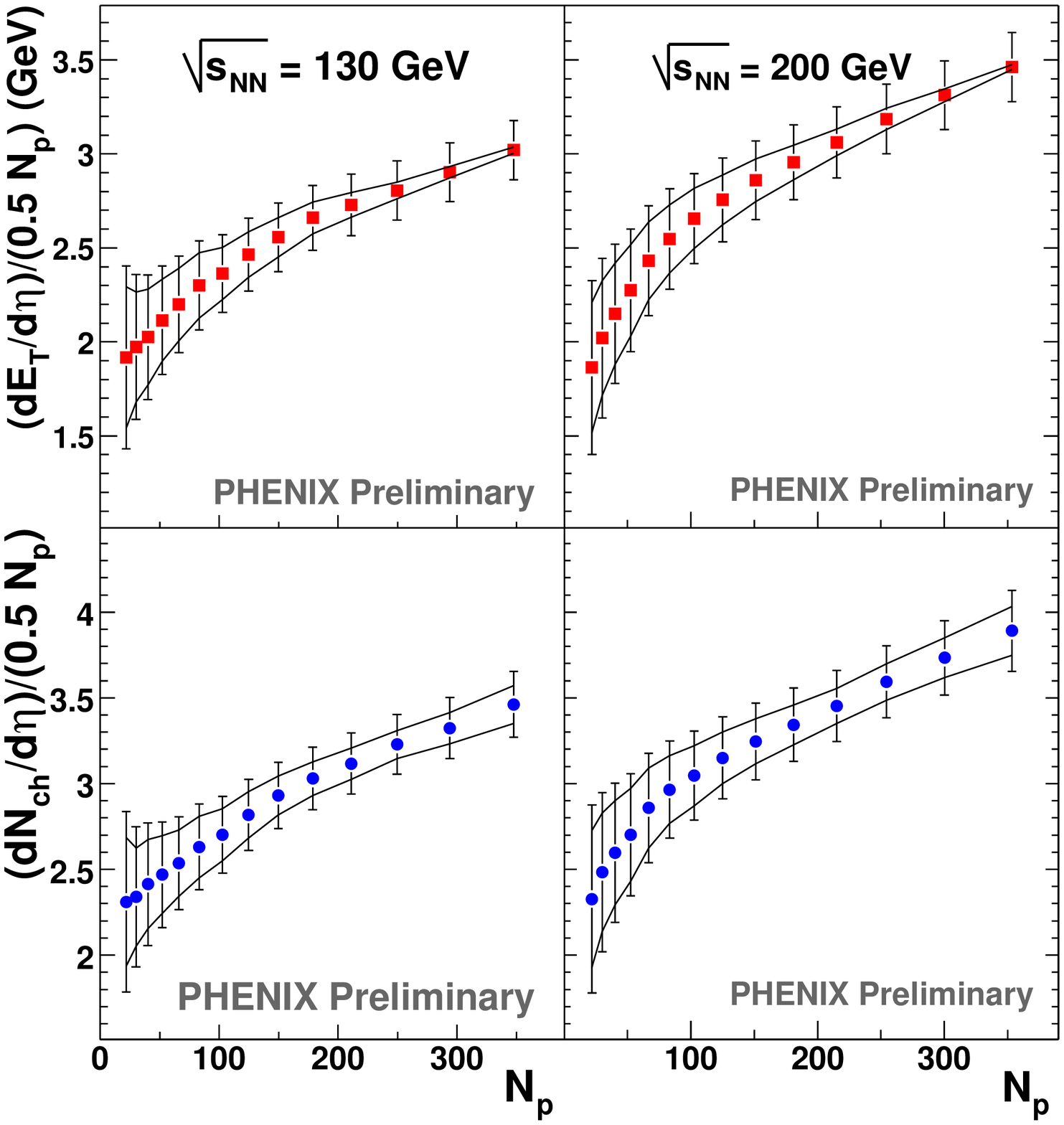}}
\caption{ $dE_{T}/d\eta$ (top panels) and 
$dN_{ch}/d\eta$ (bottom panels) per pair of participants versus $N_{p}$ 
measured at $\sqrt{s_{_{NN}}}=130$~GeV (left panels) 
and $\sqrt{s_{_{NN}}}=200$~GeV (right panels); 
The lines represent the effect of the $\pm1\sigma$ centrality-dependent 
systematic errors, the error bars are the total systematic errors.
}
\end{minipage}
\hspace{\fill}
\begin{minipage}[t]{50mm}
\centerline{\includegraphics[height=75mm]{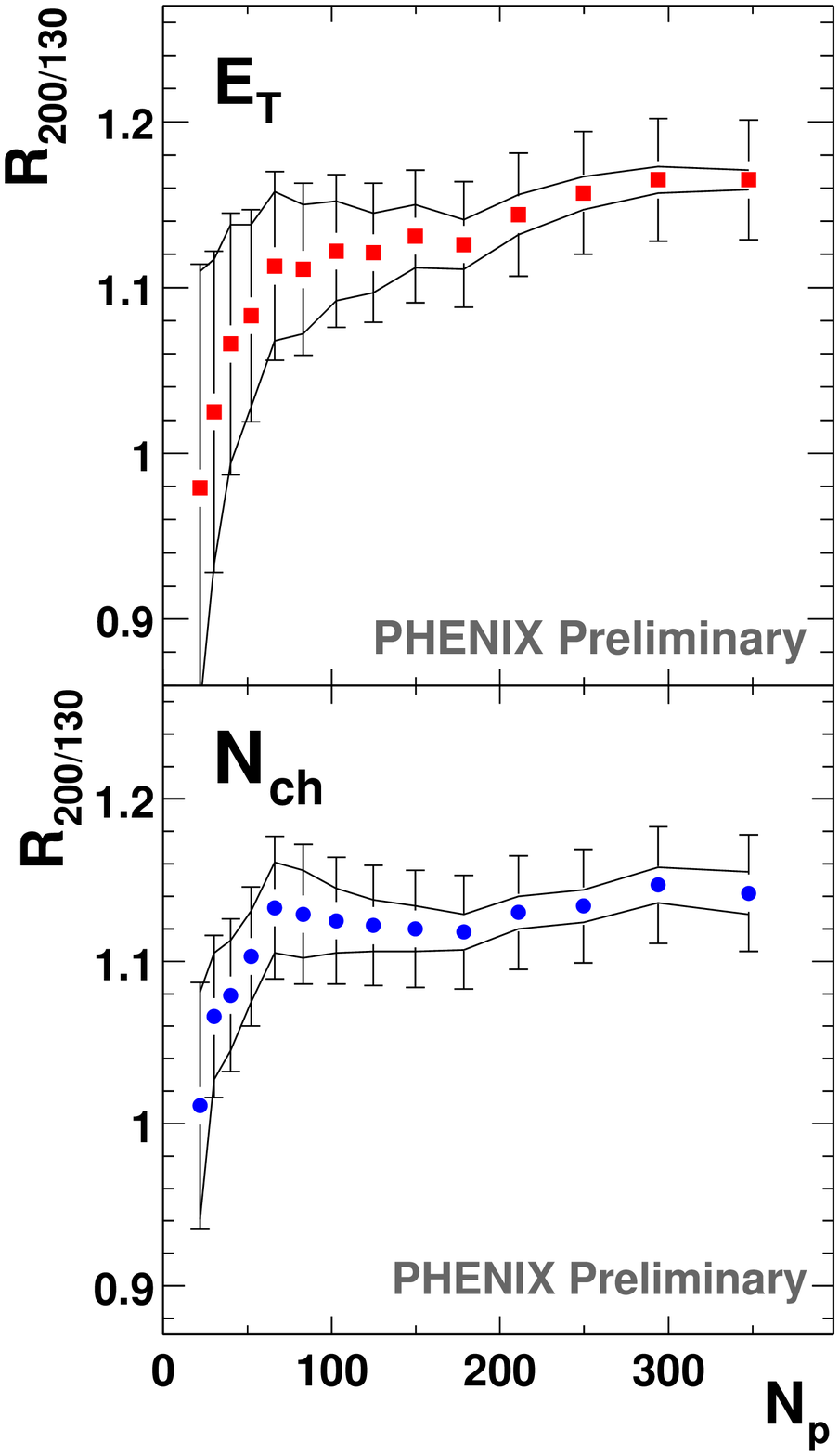}}
\caption{ $R_{200/130}$ for $dE_{T}/d\eta$ (top) and $dN_{ch}/d\eta$ (bottom)
versus centrality; $N_{p}$ is taken from data at 
$\sqrt{s_{_{NN}}}=200$~GeV; 
for the explanation of error representation, see the caption of Fig. 1.
}
\end{minipage}
\end{figure}

Fig.~1 shows the centrality dependence of $dE_{T}/d\eta$ and $dN_{ch}/d\eta$
per participant pair measured at $\sqrt{s_{_{NN}}}=130$~GeV and 200~GeV. 
Both values show a steady rise with $N_{p}$. 

The ratio of the pseudorapidity densities measured at 
$\sqrt{s_{_{NN}}}=130$~GeV and 200~GeV ($R_{200/130}$) 
for each centrality bin, corresponding to 5\% of the nuclear 
interaction cross section, is shown in Fig.~2. 
The centrality dependence of the ratios is consistent with 
a constant from the most central to semiperipheral collisions. 
For $N_{p}<70$, ratios tend to drop. 

\begin{figure}[t]
\begin{center}
\psfig{figure=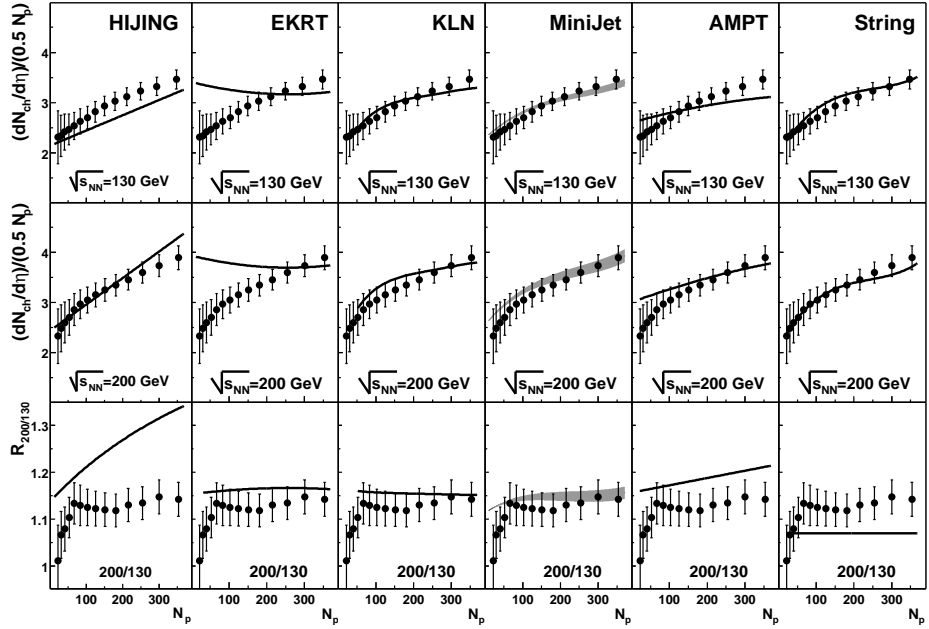,width=12.2cm}
\end{center}
\caption{The comparison of $dN_{ch}/d\eta$ at $\sqrt{s_{_{NN}}}=130$~GeV 
(top panels) and 200~GeV (middle panels) along with $R_{200/130}$ 
(bottom panels) as a function of centrality to the 
following models (solid lines): HIJING \cite{hijing}, EKRT \cite{ekrt}, 
KLN \cite{kln}, mini-jet \cite{minijet} (shaded area), AMPT \cite{ampt} 
and String Fusion model \cite{str}.}
\label{fig:model}
\end{figure}

Fig.~3 shows the comparison of our results to different model predictions. 
A powerful test for theoretical models 
is the comparison to the measured ratios $R_{200/130}$, 
since many systematic errors in experimental measurements cancel out. 
The increase of $dN_{ch}/d\eta$ with centrality is in contrast to the 
predictions of the EKRT model \cite{ekrt}. HIJING  \cite{hijing} is in 
qualitative agreement with this scenario, however the strong centrality 
dependence of the ratio $R_{200/130}$ predicted by HIJING is excluded 
by the data. Demonstrating a very good 
agreement with PHENIX measurements at $\sqrt{s_{_{NN}}}=130$~GeV 
and 200~GeV, String Fusion model  \cite{str} seems to underestimate 
the $R_{200/130}$. 
Our experimental results are well 
described by two-component mini-jet \cite{minijet} 
and high energy QCD gluon saturation \cite{kln} model calculations 
from central to semiperipheral events. The PHENIX measurements of the
ratio $R_{200/130}$ dropping at $N_{p}<70$ might set a centrality limit for 
gluon saturation model application, which predicts a near constant 
behavior of $R_{200/130}$ vs $N_{p}$.

\begin{figure}[t]
\begin{center}
\psfig{figure=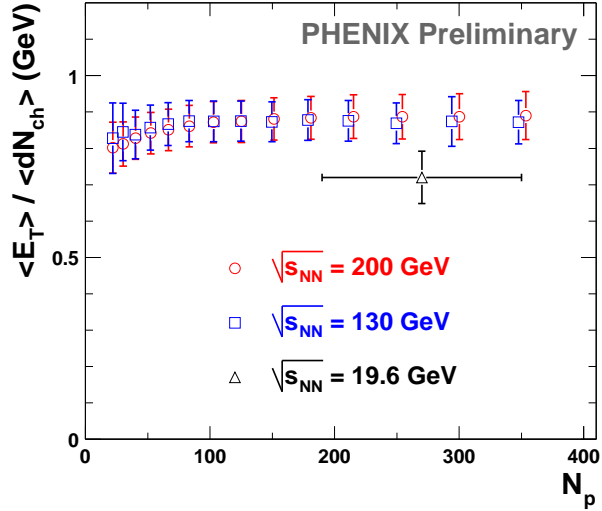,height=7cm}
\end{center}
\vspace*{-0.5cm}
\caption[]{$dE_{T}/d\eta|_{\eta=0}$ / $dN_{ch}/d\eta|_{\eta=0}$ versus 
$N_{p}$}
\label{fig1}
\end{figure}

$E_{T}$ and $N_{ch}$ behave in a very similar manner such that the mean 
$E_{T}$ per charged particle remains unchanged over a broad range of 
centralities (see Fig.~4). Fig.~4 also presents our preliminary result 
obtained at $\sqrt{s_{_{NN}}}=19.6$~GeV. Since we haven't yet finalized the 
centrality determination in $\sqrt{s_{_{NN}}}=19.6$~GeV data, 
we present our result with one point with large uncertainty in $N_{p}$ 
(this error cancels out in the ratio $<E_{T}>/<N_{ch}>$).

\section{Conclusions}\label{concl}

We presented our results for $N_{ch}$ and $E_{T}$ in a broad range of 
centralities corresponding to $N_{p}$ varying from 22 to 350. 
Our results were compared to the various
theoretical model calculations. Two-component mini-jet \cite{minijet} 
and high energy QCD gluon saturation \cite{kln} model calculations 
describe our data reasonably well. The ratio $R_{200/130}$ of charged 
particle multiplicities measured at $\sqrt{s_{_{NN}}}=200$~GeV and 
130~GeV tends to drop at $N_{p}<70$. This implies a centrality limit for 
gluon saturation model application, which predicts a near constant 
behavior of $R_{200/130}$. 

The ratio of the mean $E_{T}$ and $N_{ch}$ stays near constant over the  
broad range of centralities. It also doesn't change from 
$\sqrt{s_{_{NN}}}=130$~GeV to 200~GeV within our systematic errors. 
However, the $<E_{T}>/<N_{ch}>$ 
measured at $\sqrt{s_{_{NN}}}=19.6$~GeV was about 20\% below that 
obtained at higher $\sqrt{s_{_{NN}}}$. This is consistent with 
mean $p_{t}$ decrease from RHIC $\sqrt{s_{_{NN}}}=130$~GeV and 
200~GeV \cite{prl_h,qm02_h}
to SPS $\sqrt{s_{_{NN}}}=17.2$~GeV \cite{na49}.

\section*{Acknowledgements}
We thank the staff of the RHIC project, Collider-Accelerator, and Physics
Departments at BNL and the staff of PHENIX participating institutions for
their vital contributions. 
We acknowledge support from the Department of
Energy and NSF (U.S.A.), MEXT and JSPS (Japan), RAS, RMAE, and RMS
(Russia), BMBF, DAAD, and AvH (Germany), VR and the Wallenberg
Foundation (Sweden), MIST and NSERC (Canada), CNPq and FAPESP (Brazil),
IN2P3/CNRS (France), DAE and DST (India), LG-YF, KRF and KOSEF (Korea),
and the US-Israel Binational Science Foundation.

\vfill\eject
\end{document}